\documentclass[journal]{vgtc}                

\usepackage{mathptmx}
\usepackage{graphicx}
\usepackage{times}
\usepackage{float}
\usepackage{color}
\usepackage[normalem]{ulem}

\onlineid{309}

\vgtccategory{Research}

\vgtcinsertpkg

\title{Jacob's Ladder: The User Implications of Leveraging Graph Pivots}

\author{Alex Bigelow, Megan Monroe}
\authorfooter{
\item
 Alex Bigelow is with the University of Utah. E-mail: abigelow@cs.utah.edu.
\item
 Megan Monroe is with Tufts University. E-mail: mmonroe@cs.tufts.edu.
}

\shortauthortitle{Bigelow \MakeLowercase{\textit{et al.}}: The Intuitive Power of Graph Pivots For User Exploration and Adaptive Data Abstraction}

\abstract{This paper reports on a simple visual technique that boils extracting a subgraph down to two operations---pivots and filters---that is agnostic to both the data abstraction, and its visual complexity scales independent of the size of the graph. The system's design, as well as its qualitative evaluation with users, clarifies exactly when and how the user's intent in a series of pivots is ambiguous---and, more usefully, when it is not. Reflections on our results show how, in the event of an ambiguous case, this innately practical operation could be further extended into ``smart pivots'' that anticipate the user's intent beyond the current step. They also reveal ways that a series of graph pivots can expose the semantics of the data from the user's perspective, and how this information could be leveraged to create adaptive data abstractions that do not rely as heavily on a system designer to create a comprehensive abstraction that anticipates all the user's tasks.}


\CCScatlist{ 
 \CCScat{K.6.1}{Management of Computing and Information Systems}%
{Project and People Management}{Life Cycle};
 \CCScat{K.7.m}{The Computing Profession}{Miscellaneous}{Ethics}
}

\begin{document}


\firstsection{Introduction}

\maketitle


Graph-based data systems are everywhere. Once thought of as a fallback option for data that couldn't be finagled into a relational database, graphs are now emerging as the data format of choice, not only for overtly networked systems, such as social networks and citation networks, but also for biological systems, traffic patterns, and all of human knowledge~\cite{singhal_introducing_2012,ferrucci_ibm's_2011}.

For the domain experts who will ultimately be using this data, however, graph databases offer only a new spin on a classic conundrum: how to answer new and evolving questions. Obviously there are countless ways to explore graph data programmatically, but command-based queries often exceed the technical capabilities of end users. Conversely, reporting tools can provide answers to a predetermined set of frequently asked questions, but this relies on a technical expert to foresee and interpret the users' needs.

This latter influence, in fact, is nearly impossible to erase since it is a technical expert who must impose the initial data abstraction that will dictate how all subsequent queries will be executed. This choice of abstraction, which can be highly subjective, crucially determines how the data can be used. An ill-informed choice can dramatically reduce the efficiency and accessibility of the data for the users' most high-value tasks. It can preclude certain visualization and exploration tools from being used at all.

The ultimate goal of this work is to identify first steps towards severing the dependence of a graph's utility on its initial data abstraction. To do this, we focus on a graph-based operation known as a ``pivot.'' The pivot allows users to evaluate one set of nodes in the context of some subset of its neighbors. It offers the unique advantages of its visual complexity being agnostic to the graph's size, and its simplicity making it compatible with.

We present an overview visual technique, dubbed Jacob's Ladder, which allows users to traverse, query, and extract sub-sections of a graph using only chained sequences of pivots and filters. We report on how the strengths and weaknesses of this technique's design influence users' ability to grok the underlying data abstraction. Using this tool, we are able to observe where and how ambiguity can arise in a series of pivots. We propose ``smart pivot'' heuristics as a means of overcoming these natural ambiguities. Finally, we discuss the potential of graph pivots in exposing inconsistencies between the data abstraction and the users' needs. We outline examples for how these pivots could inform an \textbf{adaptive abstraction}, in which a system reshapes its schema on the fly to become more semantically relevant and efficient as questions are asked, rather than rely on a technician's \textit{a priori} intuition about what future users' questions might be.

\section{The Pivot}
As shown in Figure~\ref{fig:traversal}, we define a \textbf{pivot} as an operation in which a user navigates from a set of \textbf{seed} nodes $S$ to another set of \textbf{target} nodes $T$, in which every target node $t\in T$ has a connection to at least one seed node $s \in S$. Note that the sets of seed and target nodes need not have any internal structural relationship; $S$ and $T$ may be arbitrarily large, and the members of each set may be entirely disconnected from members of the same set.

This operation can be chained together, with the target nodes $T_0$ from the previous step serving as the seed nodes in the current step: $T_0 = S_1$. For example, in a simple social network of friends, the ``friends of friends'' for any node can be found by performing two pivots. While the pivot, by itself, creates a fairly simplistic fan-out effect, it is considerably more expressive with these common extensions:

\begin{figure*}[t]
\begin{center}

\includegraphics[width=\textwidth]{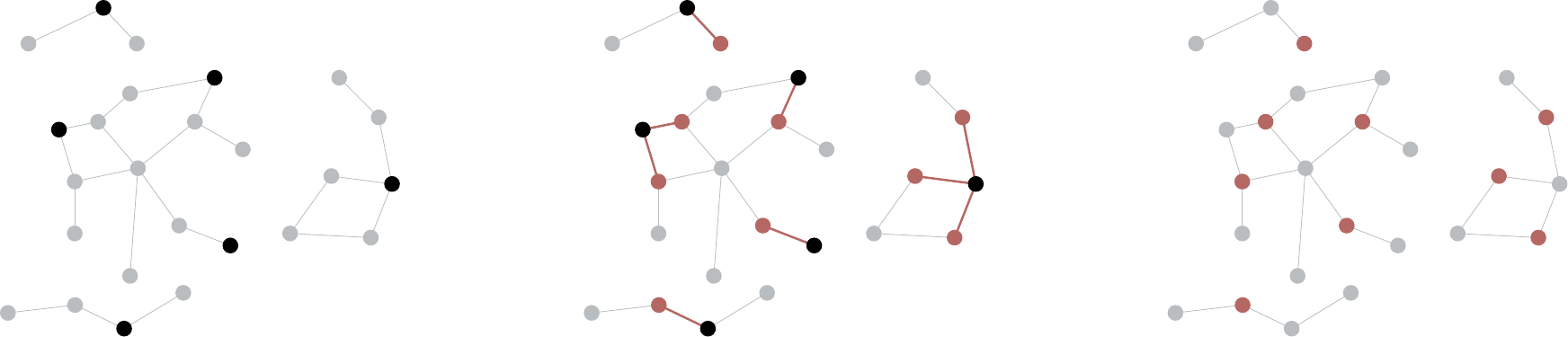}

\caption{The basic pivot: On the left, the dark set of seed nodes are selected. The selection then swings out to a subset of neighboring target nodes (middle, red), resulting in a new set of seed nodes (right).}
\vspace{-0.6cm}
\label{fig:traversal}
\end{center}
\end{figure*}

\subsection{Categorical Pivoting}
A pivot does not necessarily need to swing out to all of the connected neighbors of the seed set. When a graph is heterogeneous, consisting of multiple types of nodes and edges, a pivot can swing out to only nodes of a certain type or along edges of a certain type or both. For example, in the data system for a large hospital, a doctor, Alice, might want to find out which other doctors her patients are seeing. By first finding herself in the data system (for example, $D_0 = \{ Alice \}$), she can pivot out to all of her patients (for example, $P = \{ Bob, Carol \}$), and then pivot back to all of the doctors associated with those patients (for example, $D_1 = \{ Alice, Dave, Eve \}$).

\subsection{Filtering}
After any pivot, if a graph is multivariate, users may want to filter the subgraph of seed and neighbor nodes before the next pivot is performed. For example, instead of finding the other doctors that all of her patients are seeing, maybe Alice only needs to find the other doctors of her female patients. In this case, the set of patient nodes can be filtered down to just the female patients (for example, $P' = \{ Carol \}$) before performing the second pivot back to doctors (for example, $D_1 = \{ Alice, Eve \}$). This filtering can be based on node attributes, edge attributes, the number of incoming or outgoing edges, or any other metric that can be computed against the subgraph of seed and neighbor nodes. Filtering, as well as categorical pivoting, make it possible to perform multiple consecutive pivots without continually increasing the number of nodes involved in each pivot, achieving a fan-in effect.

For our purposes, we will describe \textbf{direct} filters as those performed directly on a set of nodes, such as filtering patients nodes by their sex attribute. We will describe \textbf{connective} filters as those that indirectly filter a different set of nodes, such as the second set of doctors ($D_1$) having been being filtered indirectly by their patients' sex.

\section{Related Work}
Pivots have made both direct and indirect appearances across the graph visualization literature. In using the term ``pivot,'' we refer to it in the sense of traversing an existing graph, from one set of nodes to another~\cite{kang_exploring_2006,wattenberg_visual_2006,stasko_jigsaw:_2007,vanham_"search_2009,dunne_graphtrail:_2012,dork_pivotpaths:_2012,ghani_multinode-explorer:_2012}, rather than toggling between node and edge interpretations~\cite{renoust_detangler:_2015}, or aggregating node attributes in the process of modeling a graph~\cite{liu_ploceus:_2014}. While pivots have been identified and used in the past---we do not claim the identification of pivots as a contribution---their usage is typically limited to an initial seed node set of size one; chaining pivots together is often not supported; pivots are used to support specific tasks on specific data abstractions; and/or pivots are integrated as part of a broader system that does not give an opportunity to study them in isolation. This work explores the power and effects of graph pivots in general, ignoring any particular abstraction.

In terms of Lee \textit{et al.}'s Graph Task Taxonomy~\cite{lee_task_2006}, a graph pivot falls into three of the four identified categories. It is a \textbf{topology-based} operation in that it starts by identifying the neighbors of the seed nodes. It is an \textbf{attribute-based} operation in that it filters the neighbors to the set of target nodes. Finally, it is fundamentally a \textbf{browsing operation}, in that it traverses a set of $n$ paths through the graph simultaneously. As pivots are essentially an aggregate form of traversal, there is no comparison to be made to traditional instance-based techniques, such as node-link diagrams or adjacency matrices. Rather, the technique that we demonstrate could be used in conjunction with standard instance-based techniques in a linked view system. Testing our technique in isolation allows us to reflect on whether and why additional views may be necessary.

\section{Why the Pivot?}
There are three reasons why the pivot stands out as a potential linchpin of usable graph exploration:

\subsection{Manageable Subgraphs}
As graph data stores become larger and increasingly complex, the assumption that the graph can be loaded into memory and visualized in its entirety will eventually stumble. Thus, in isolation, systems such as Gephi~\cite{bastian_gephi_2009}, g-Miner~\cite{cao_g-miner:_2015}, Tulip~\cite{auber_tulip_2004}, and a host of others~\cite{shannon_cytoscape:_2003,bonsignore_first_2009,bezerianos_graphdice:_2010} that rely on a holistic display of the graph, will fail to scale.

In contrast, the pivot provides a consistent and easy-to-interpret means of displaying a partial view of the underlying graph. In order to perform consecutive pivots, users only need to see their current subgraph of seed and target nodes, and the options for where they can pivot next. As we show in this paper, novice users can extract and understand meaningful subsets of a graph by employing only pivots and filters, even though the topology of individual nodes and edges remains hidden. The pivot can be executed and visualized without having to account for the size and complexity of the entire graph.

\subsection{Coverage}
Before users can analyze data, they must first be able to isolate the data that is relevant to their questions. This can be a steep challenge when users do not have flexible access to the underlying data system. This work was motivated, in part, by a series of interviews with a group of bank employees who regularly interacted with a large reporting tool system. Users expressed consistent frustration with not being able to investigate connections between elements that were, in fact, connected in the underlying data. The phrase we heard over and over again was, ``I can't get from \underline{\hspace{1cm}} to \underline{\hspace{1cm}}.''

The pivot operation addresses this difficulty by allowing movement to take place across any existing connections in the underlying graph. So long as the graph is connected, pivoting allows users to navigate between any two nodes in the system. This navigation might not precisely represent the intent of the user's ultimate objective, but as we will discuss in subsequent sections, it can significantly narrow down the space of what that objective might be.

\subsection{Abstraction Agnostic}
It is easy to underestimate the subjectivity of a graph's data abstraction~\cite{munzner_nested_2009}. Decisions must be made about what will be a node, what will be an edge, and what will be the attributes of those nodes and edges. A city, for example, could be easily viewed as an attribute of a university node (i.e. the city in which that university is located). However, it might make more sense for each city to be its own node in the graph, and for the location of a university to be represented by an edge to that city node. Flipping the notion of nodes and edges entirely can also produce a more intuitive graph~\cite{nielsen_abyss-explorer:_2009}. The overall utility of a graph can depend heavily on how well the data abstraction matches the queries that will ultimately be run against it. We refer to these arbitrary abstraction decisions as the \textbf{schema} of the graph, including: what is a node; what is an edge; what node or edge types exist; whether a graph is undirected, directed, or mixed; whether parallel edges are allowed; whether structures such as supernodes or hyperedges are supported; and whether nodes and/or edges are multivariate.

The advantage of the pivot is that it is simple enough to work on any graph schema, as long as one has been identified. However a graph is represented internally---whether an adjacency matrix, a node-link list, or modeled from a relational database~\cite{heer_orion:_2011,liu_ploceus:_2014}---the concept of a pivot is still valid. This universal applicability can be contrasted with techniques that require restrictive assumptions about what the underlying data will be, and how it will be organized~\cite{kang_exploring_2006,stasko_jigsaw:_2007,dork_pivotpaths:_2012}. Unlike pivots, techniques that require schema definitions beyond simply having nodes and edges are immediately ruled out when a dataset doesn't match the needed abstraction.

Pivots also have the potential to reveal the semantics of the user's tasks. Consider the hospital example: if our doctors need to find the other doctors that their patients are seeing, they can find themselves, pivot to patient nodes, and then pivot back to doctor nodes. The series of pivots explicitly encodes the semantics of the doctor's intent in a very simple way that could be collected to better understand users' needs and improve the underlying data abstraction. We discuss two specific ways this could happen in more detail in Section~\ref{sec:adaptiveAbstraction}.

\section{Why Not the Pivot?}
\begin{figure}[t!]
\begin{center}

\includegraphics[width=\columnwidth]{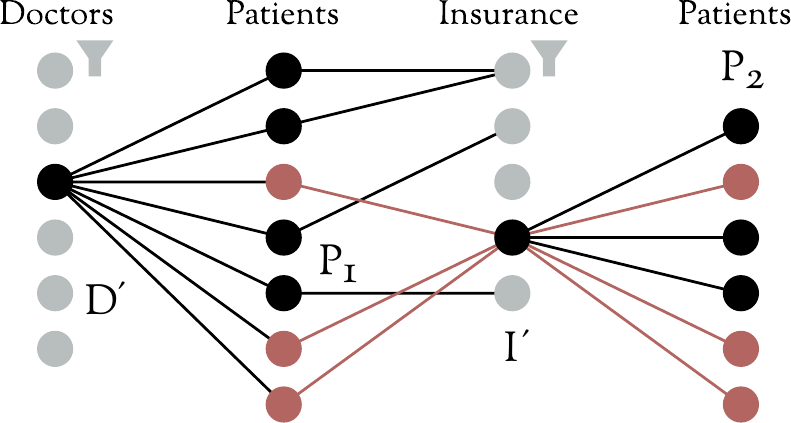}

\caption{To find patients of a given doctor that are covered by a certain insurance provider, the user starts by filtering the doctor nodes down to a single doctor ($D'$). The user then pivots to patients ($P_1$), then to insurance providers ($I'$, where another filter is applied), then back to patients ($P_2$). However, when the user pivots back to patients, the pivot returns all of the patients with the specified insurance provider, but not necessarily patients of the original doctor (in red).}
\vspace{-0.5cm}
\label{fig:medical}
\end{center}
\end{figure}

The obvious drawback of formulating meaningful queries or explorations by chaining together a series of pivots is that each pivot operation is atomic. A single pivot sees only the seed nodes and their immediate neighbors, not the series of pivots that led up to that point. The ramifications of this limit can be illustrated by the following scenario: consider a doctor ($D'$) who would like to know what kinds of treatments they have prescribed to patients ($P_0$) with a particular insurance provider ($I'$). The resulting sequence of pivots, shown in Figure~\ref{fig:medical}, might start with doctors locating themselves in the data system, pivoting out to their patients, then pivoting out to the insurance providers of those patients, and filtering that list down to the provider of interest. But now our doctor has a problem. Pivoting back to patients ($P_1$) will yield a list of all the patients who have that insurance provider, not necessarily \emph{that} doctor's patients who have that insurance provider. That next pivot doesn't inherently understand that the pool of patients was already narrowed down and, as a result, the pivot sequence starts to diverge from the intent of the query. This difficulty can be reproduced in a system like GraphTrail~\cite{dunne_graphtrail:_2012}, which only looks at a single pivot at time.

This speaks to one of our main contributions: is it possible to make these pivots smarter, so that their meaning is always unambiguous? The benefits of such an improvement are twofold. Users would, of course, have a more expressive, powerful way to query and navigate a database. Additionally, clearing up this ambiguity supports our final major contribution: the simple, unambiguous nature of a series of pivots will allow researchers and systems to collect data that directly exposes what users are actually looking for.

\section{Evaluating Pivots}
The critical weakness of pivots, as we have discussed, lies in the ambiguity that arises as the user traverses deeper into the graph with a series of pivots. Where does this ambiguity come from, and what could a user do to help clarify it?

To better understand the translation between real-world questions and sequences of graph pivots, we implemented the pivot operation as a web-based front-end to a Titan graph database, using the Gremlin query language. Although traversal languages such as Gremlin are well-suited to computing pivots in our case, we attempt to focus on how users understand pivots, rather than how to compute pivots efficiently---these technology choices may not be appropriate or efficient for every graph data abstraction.

\subsection{Interface Design}
\begin{figure}[t!]
\begin{center}

\includegraphics[width=\linewidth]{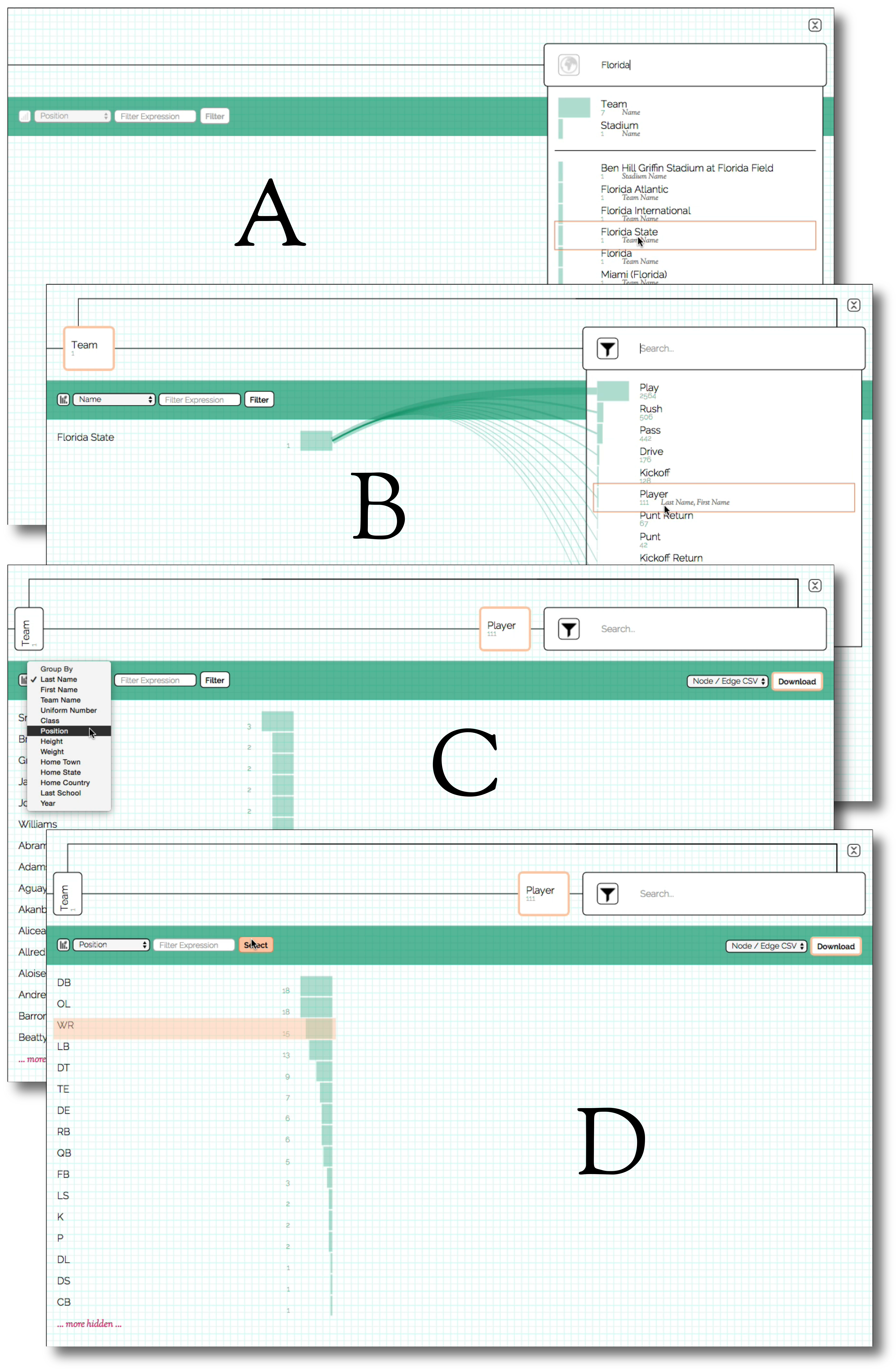}

\caption{Jacob's Ladder allows users to pivot from one category of nodes to another. A search box (A) shows search matches in the menu below. Matching nodes can be selected in aggregate, based on node type (``Team'' or ``Stadium'' above the line), or individually based on value (below the line). Once a set of nodes has been selected, it is displayed as a histogram on the left of the search field (B). Subsequent searches are limited to the set of nodes that are connected to the previous selection, with line thickness encoding potential connections. The histogram supports regrouping and sorting (C), as well as selecting and filtering nodes (D) based on node attributes. The series of actions depicted are as follows: A) Florida State is selected, B) the user pivots to Florida State's players, C) players are grouped by position, and D) the wide receivers (``WR'') are selected.}
\label{fig:tool}
\end{center}
\end{figure}

\begin{figure}[htbp]
\begin{center}

\includegraphics[width=\linewidth]{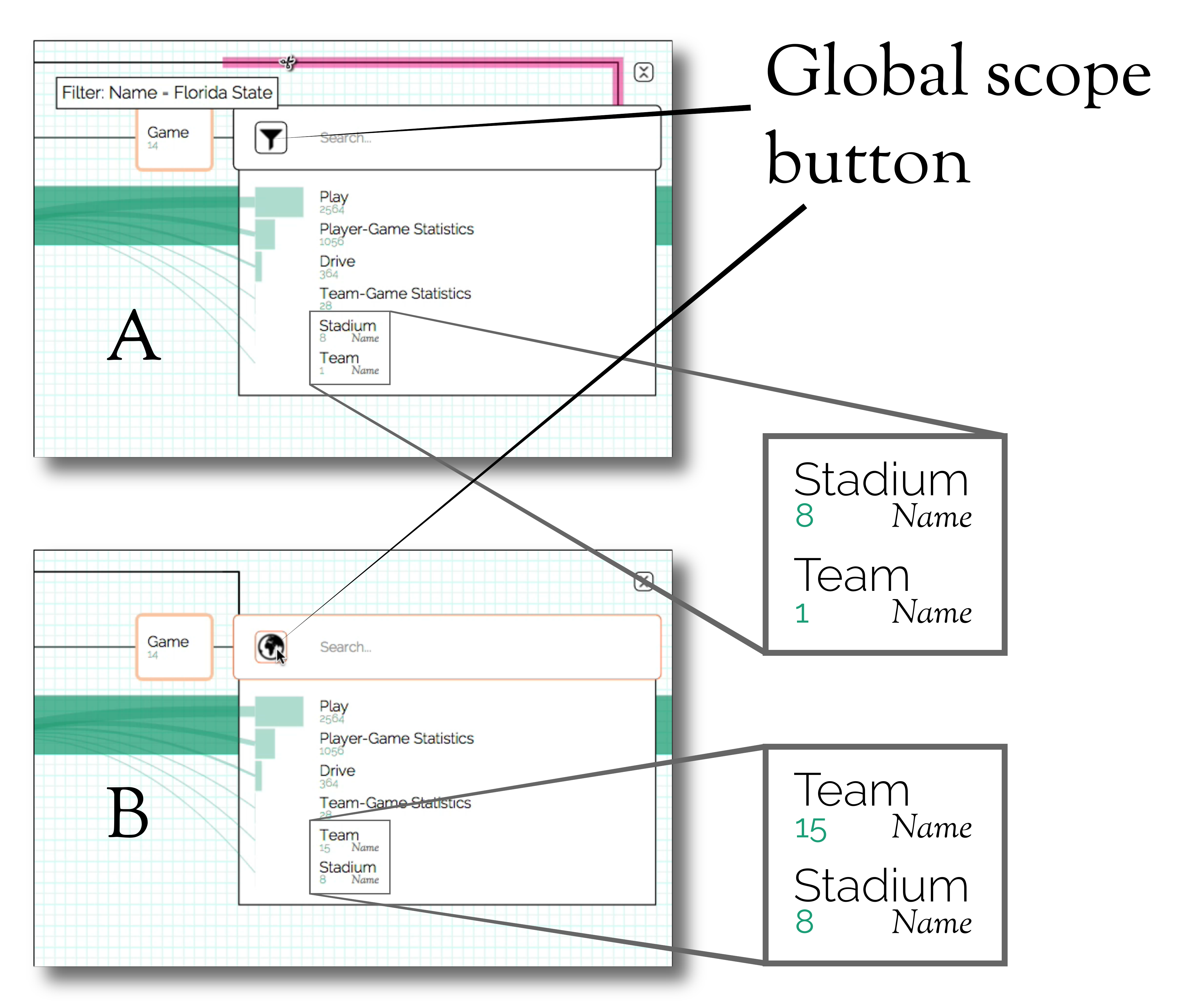}

\caption{When filters are applied to a selection of nodes, a line is placed at the top of the interface to indicate that the filter is active (A). Because the difference between fanning in and fanning out is so critical, it can be toggled in two ways: a global scope button inside the search field removes or restores all filters, or individual filters can be removed by ``snipping'' the line. Note how, in A, only one Team node can be selected, because the filter is still in place. Clicking ``Team'' will fan in. In B, because the global scope button has been clicked, the set of available Team nodes is larger; the filter has been removed. Clicking ``Team'' will fan out.}
\label{fig:globalScope}
\end{center}
\end{figure}

The resulting application, dubbed Jacob's Ladder, is shown in Figures~\ref{fig:tool} and~\ref{fig:globalScope}. It allows users to select an initial set of seed nodes, apply filters to the set, and then pivot to a new set of connected nodes. This process can be repeated as many times as needed, with a summary of previous pivots and filters represented as lines across the top of the screen for reference. At any point, users can undo a pivot, an associated filter, or clear their history of pivots and start from scratch. As this work focuses on understanding the role of previous filters in the context of subsequent pivots, Figure~\ref{fig:globalScope} shows how filters can be inspected and edited individually using filter lines, or toggled across the board using a global scope button in the search bar.

The interface is designed primarily for subgraph extraction. As a user pivots through the graph, the consecutive sets of seed nodes form a smaller, more manageable subgraph that can be downloaded as common graph formats that include the edges used in the traversal. Ideally, Jacob's Ladder should be used to extract a meaningful, manageable subgraph from a large database for closer analysis in other tools, such as Gephi~\cite{bastian_gephi_2009}---Jacob's Ladder is not designed to support low-level, per-node analysis. Visualizations of individual nodes and edges are deliberately omitted from its interface.

Because Jacob's Ladder operates at such a simple, aggregate level, it completely bypasses the scale problems of traditional graph visualization systems. The required screen real estate is a function of the various types of nodes and edges in the schema of the graph, not the actual number of nodes and edges. Consequently, there is no visual limitation with respect to the actual size of the graph.

\subsection{Lab Tests and Design Adjustments}
The limited scope of Jacob's Ladder presents an opportunity to study graph pivots in relative isolation. Over the course of three months, we loaded Jacob's Ladder with a wide range of graph datasets, from IMDB's movie graph to financial and medical data, and tested where and how ambiguities arise in the pivoting process.

We can further simplify our discussion of pivots if we treat edges as distinct entities---our experience designing Jacob's Ladder itself yielded this insight. Where relevant, to allow a simpler interface, the tool reinterprets any edges as interleaving nodes. For example, if a relationship edge in a social network has attributes, it would be replaced with an edge, a node containing those attributes, and another edge. Early prototypes of the system maintained a distinction between the two---however, the redundancy became obvious very quickly. For the sake of simplicity, we chose to avoid additional UI elements that differentiate between data on nodes and edges.

As we designed Jacob's Ladder and used it to explore these datasets in the lab, we came to develop a prediction that \textbf{ambiguity only arises in a series of pivots when the series includes both filters and cycles.}

\subsection{Qualitative Evaluation}
To learn how users understand graph pivots, whether they are useful, where ambiguity arises, and how pivots reveal a user's semantic understanding of a graph, we conducted an informal, qualitative study of users. Because we had developed some initial predictions, we were careful to design our experiment to evaluate those predictions explicitly. We were also careful to watch for trends that we did not anticipate, including unexpected or surprising behavior.

\subsubsection{Participants}
Initially, this system was developed for internal use within a large financial institution, and its use in a hospital database was also anticipated. Unfortunately, due to confidential data and legal complexities, we were not able to gain access to real users in or out of their native work environment.

\begin{figure*}[htbp]
\begin{center}

\includegraphics[width=\textwidth]{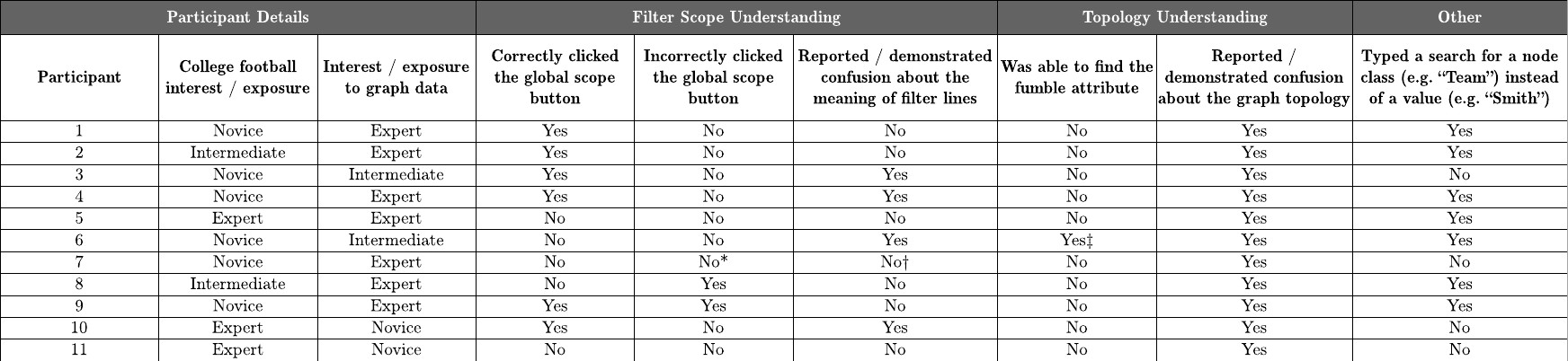}

\caption{This table shows details about each of the eleven participants, including their relative expertise and suggestive indicators that emerged as the study progressed. Participants are classified as ``Novices'' when they self-reported little to no understanding or prior experience, ``Intermediate'' when they reported or demonstrated some familiarity, but no strong interest or experience, and ``Expert'' when they reported or demonstrated strong interest or experience. $\ast$ This participant briefly clicked the button at an inappropriate point, but quickly reverted the decision. $\dagger$ This participant specifically asked about the filter lines, so they were given an explanation. $\ddagger$ Technically, this participant found the ``fumble return'' attribute---a different attribute of the Player-Game Statistics node type. Structurally, this is equivalent.}
\vspace{-0.5cm}
\label{fig:participants}
\end{center}
\end{figure*}

Consequently, we selected a publicly available NCAA American College Football dataset. This dataset was interpreted as a graph with many node types, such as Players, Teams, Games, Stadiums, Conferences, etc. As shown in Figure~\ref{fig:participants}, a diverse range of participants were selected, from graduate students that have experience with graph data but minimal knowledge of football, to passionate football fans with little to no graph exposure. Each was asked to self-report their understanding or expertise with regard to graph data and American football.

\subsubsection{Hypotheses and Tasks}
The interface of Jacob's Ladder provided an opportunity to assess both the power and limitations of graph pivots. Specifically, we designed tasks that address the following research questions:
\begin{enumerate}
\item Can the graph pivot enable technical and domain novices to extract meaningful subsets of a large graph, even when traditional instance-level visualizations are not included?
\item How does the technique obscure the topology of the database?
\item Does the user understand the scope of the next pivot? Is the interface sufficient to resolve ambiguous cases?
\end{enumerate}
The corresponding tasks are:
\begin{enumerate}
\item With minimal introduction, observe whether users can select all the quarterbacks on a specific team (users must filter, then pivot,
then filter).
\item Observe whether users can anticipate where to find the ``fumble'' attribute without help (filter, pivot, connective filter, pivot back).
\item Given a specific team, observe whether users can select the set of teams that the seed team beat. This task requires the user to either remove the initial filter, or toggle the global scope button (filter, pivot, filter, toggle scope, pivot back).
\end{enumerate}

It is important to note that these tasks were designed to aggressively discover the limitations of our technique, rather than merely serve as existence proofs of where it succeeds~\cite{greenberg_usability_2008}. Consequently, we focus on these limitations in discussing our observations, as they form the seeds for reflection in Section~\ref{sec:discussion}.

\subsubsection{Experiment}
Each 30-minute session involved the participant and the researcher seated at mirrored displays, each with a mouse and keyboard. In addition to the researcher`s notes, screen capture software was used to record the user's actions and voice. Where necessary, participants were first given a brief introduction to the dataset, including explanations about college football and/or graph data. Participants were then given a 5-minute introduction to the tool, including two brief demonstrations of the system, similar to tasks 2 and 3 that the participants would later be given. As we were particularly interested in understanding whether users could decipher the scope of their applied filters on their own, only the function of the global scope button was explained and demonstrated; the filter lines were ignored. Next, participants were given the three tasks in order. Finally, users were given time to explore the data freely, and comments, questions, and discussion were encouraged. As we were particularly interested in understanding whether users could decipher the scope of their applied filters on their own, only the function of the global scope button was explained and demonstrated; the filter lines were ignored.

\subsubsection{Task 1 Observations}
All participants were able to accomplish the initial filter and pivot in Task 1 with ease. Interestingly, while most users were able to perform the final filter without difficulty, many users were not aware that they had already successfully completed Task 1.

Participants navigated from Team nodes to Player nodes, and the interface initially displayed the principally descriptive attribute of the nodes they had selected: in this case, player names. Users would switch the histogram to group players by their position attribute, and then filter players by selecting the ``QB,'' or quarterback position. At this point, users had technically succeeded in selecting the quarterback player nodes, but because the histogram only displayed one ``QB'' bin, they often were not aware that they were finished until they switched back to the player name attribute.

\subsubsection{Task 2 Observations}
In Task 2, participants were asked to find the set of players on a team of their choice that had fumbled the ball at some point in the season. This question was difficult for all participants to perform because it required traversing from Team nodes to Player nodes, and then to Player-Game Statistics nodes. No ``fumble'' attribute was directly visible from the Player nodes. As such, only one participant was able to come close to successfully navigating to this set.

\subsubsection{Task 3 Observations}
The third task was to identify the set of teams that a team of their choice beat. This was an opportunity to observe whether users understood the scope of the filters. We specifically tracked whether participants clicked the global scope button at the correct point in the task.

Though a very similar demonstration was shown to each participant at the beginning of the study, they displayed mixed results in their success. The fact that filters were still in place as they pivoted back to a previous node type (Team $\rightarrow$ Game $\rightarrow$ Team) appeared to be somewhat unintuitive.

\subsubsection{Incidental Observations}
While the study was somewhat controlled by the tasks issued to the participants, we were careful to observe whether additional patterns surfaced.

We observed some confusion between the filter functionality of the tool and the pivot functionality. Participants would sometimes go to the search box when they meant to filter, or to the filter controls when they meant to pivot. This reveals a design flaw in the Jacob's Ladder interface: the search field technically applies a filter, as do the more traditional filter controls. Applying filters in multiple locations in the interface caused some confusion.

Another unexpected pattern that we observed was that participants would often enter a node class name, such as ``Team'' in the search box instead of attribute values. Because the system only expects attribute queries in the search field, it would try to find node attributes that match ``Team'' instead of finding nodes by class name. ``Team'' nodes would subsequently disappear from the menu, resulting in confusion.

Finally, we were surprised by how well the participants were able to interpret the meaning of their current selection. Particularly during Task 2, participants were observed performing long chains of pivots in search of the ``fumble'' attribute. Almost all participants were cognizant of the fact that they needed to be somewhere else in the graph. Impressively, almost all participants were able to articulate the meaning of their current selection when asked, even if many pivots were involved. For example, during Task 3, Participant 5 navigated from Team (Ohio State) $\rightarrow$ Team-Game Stats (WIN) $\rightarrow$ Team (failed to remove the Ohio State filter) $\rightarrow$ Game $\rightarrow$ Team-Game Stats (WIN filter still applied, grouped by team name). When asked, he correctly interpreted the visible set as any team that had won a game that Ohio State was involved in.

\section{Discussion}
\label{sec:discussion}
Overall, our results support, with some qualifications, our hypothesis that the graph pivot is a powerful tool that can enable novice or disinterested users to extract meaningful subsets of the graph without visualizing low-level graph topology. The tests also confirmed our predictions about where ambiguity arises in a series of pivots. Finally, the tests showed that a series of pivots can expose the user's understanding of the semantics of the data in a way that could easily allow for a system to reshape its data abstraction based on user behavior.

\subsection{When Are Other Views Needed?}
Our tests confirmed Hypothesis 1, that the simple pivot operation can empower novice users to extract meaningful subsets from large graphs. Our aggregate visual technique that lists each pivot at the top of the interface circumvents the scalability issues of traditional graph visualizations by avoiding local topology altogether---we demonstrate that, for many graph data tasks, it is not necessary to render detailed node-link diagrams. Working at the aggregate level that pivots enable is often sufficient and intuitive for many graph visualization tasks.

While visualization of local topology is not necessary for many tasks, we also learned from participants' performance in Task 2 that our particular implementation of the graph pivot in Jacob's Ladder obscures the global topology of the graph---we were perhaps too minimal in its design. An even higher-level overview of the schema of the graph, such as the technique demonstrated by Van den Elzen \textit{et al.}~\cite{vandenelzen_multivariate_2014}, is still likely necessary to help the user plan how to pivot and filter toward node types and attributes of interest, especially for unfamiliar datasets.

\begin{figure*}[htbp]
\begin{center}

\includegraphics[width=\linewidth]{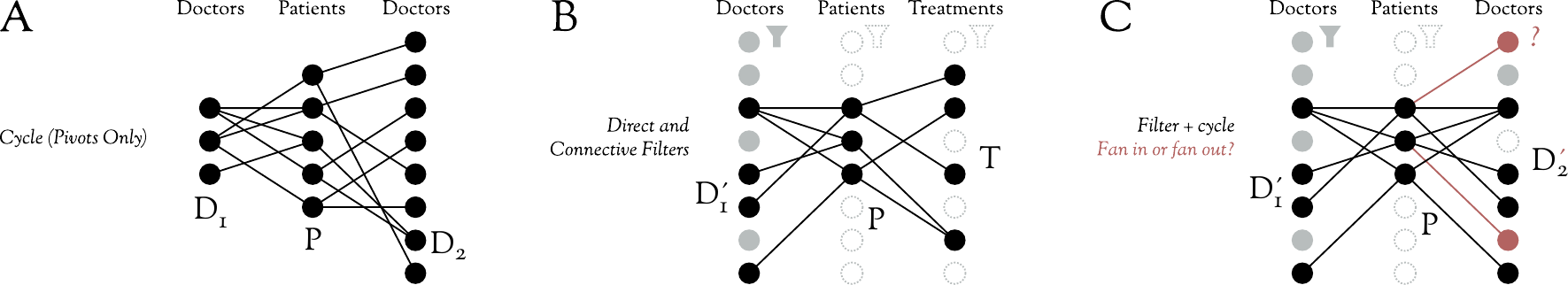}

\caption{We can see that ambiguity in a series of pivots only arises when filters and cycles occur in the same traversal; when cycles are present without filters (A), the only logical action is to fan out. When filters are present, without cycles (B), the only logical action is to keep the filter in place and fan in. However, when both are present (C), it is not clear whether to fan in or fan out: should the initial filter on the Doctor nodes be reapplied?}
\vspace{-0.5cm}
\label{fig:ambiguity}
\end{center}
\end{figure*}

\subsection{Delineating Where Ambiguity Occurs}
Task 3 confirmed our initial predictions about where ambiguity arises. As shown in Figure~\ref{fig:ambiguity}, the meaning of a user's pivot is always clear unless a cycle and a filter are encountered together.

\subsubsection{Pivots Only}
In our initial explorations of the data before the user study, the first thing we discovered was that it was impossible to create ambiguity by performing pivots without filters (Figure~\ref{fig:ambiguity}A). If no filters are enacted during a series of pivots, then the only logical outcome of the next pivot is to return all of the connected nodes of the specified category---however, pure pivots without filters may not be very useful.

\subsubsection{Pivots and Filters}
When a filter is enacted at a certain point in the pivot sequence, it manifests in two ways (Figure~\ref{fig:ambiguity}B). The first is as a direct filter against the category to which it has been applied. For example, if users want to see all of the doctors that are women in a medical database, they can group doctors $D$ by a ``gender'' attribute, and select only the group of women, resulting in a subset $D'$. This filter is applied directly to the doctor category, and depends only on an attribute of the doctor nodes.

However, when users pivot from doctors to their patients, that gender filter against the doctor category serves a dual function as a connective filter against the patient category; the resulting set of patients $P$ would likely be larger, had the filter not been applied to $D'$. This connective filter has an increasingly indirect effect on each category of nodes that is visited after the filter is applied.

Our study showed that these indirect effects were not difficult to understand. Even though users sometimes became ``stuck'' in their exploration of the football dataset after a long series of pivots and filters, they could still generally articulate the meaning of the nodes that they had arrived at, including the effects of upstream filters. Additionally, so long as no category is visited more than once after the filter has been applied, the only logical outcome is still to pivot out across all of the available connections. There is no previous interaction with that next category to suggest otherwise, and thus, no ambiguity.

\subsubsection{Pivots and Filters and Cycles}
As illustrated in Figure~\ref{fig:ambiguity}C, ambiguity arises only when a given category of nodes is visited more than once after a filter has been applied. When this occurs, the revisited category is carrying with it a set of direct filters that the user might or might not want to restrict the current pivot operation. Continuing with our previous example, where we filtered the list of doctors $D_0$ to only see women ($D_0'$), and pivoted out to patients ($P$), if we then pivot back to doctors, which doctors does the user want to see? We know that the user wants to see the doctors ($D_1$) associated with those patients , however, should the original direct filter on the ``gender'' attribute remain for this second set ($D_1'$)?

More generally, these options can be described as:

\begin{enumerate}
\item Perform the pivot operation normally, swinging out to all of the connected neighbors that match the specified category, keeping connective filter effects, but without re-applying previous direct filters (Fan-out pivot).

\item Further restrict the nodes returned by a normal pivot---retaining both connective filter effects from other categories, as well as re-applying previous direct filters on that category  (Fan-in pivot).
\end{enumerate}

\subsection{Implications For Smart Pivots}
The question then, is how to determine which of these options the user intends and, if it's the latter option, whether the user intends to retain all of the direct and connective filters that have been applied, or only a subset of them. While it is not possible to guess the exact intent of the user at every turn, we can better narrow down this problem space to isolate the exact source of the ambiguity. Our experience with Jacob's Ladder and its user tests are suggestive of heuristics to follow for intuitive behavior in ambiguous cases.

As described above, we only encounter ambiguity in the case where a user is pivoting back to a category to which they had already applied a direct filter---for example, consider the series of pivots from a filtered set of actors $A_0'$, to movies $M_0$, to directors $D$, to movies $M_1$, to actors $A_1$ (or $A_1'$, the question being whether to keep the direct filter on $A_1$). We can assume that the meaning of the first set, $A_0'$, was unambiguous when the user applied the filter to it. The interim pivots ($M_0, D, M_1$) between that point and the returning pivot to $A_1$ are therefore the source of ambiguity that we must decipher.

Jacob's Ladder itself does not implement any ``smart pivot'' heuristics---we include these heuristics as insight based on what we saw when we deliberately challenged users with questions about the data that led to both fan-out and fan-in scenarios. Users often failed to remove filters when they needed to. However, they almost never reenacted filters incorrectly. Therefore, we propose the following heuristics, and advocate for testing them formally in future work.

\subsubsection{Returning After Intermediate Filters}
The ability to enact connective filters is powerful---in the above example, we could apply a filter to directors $D'$, such as $age > 40$, whereupon the resulting traversal results in a connective-filtered set of actors \textit{of the original set} $A_0'$ that worked in films whose directors were over the age of 40. We suspect that erring on the side of fan-in---leaving the direct filter in place---will do the right thing most of the time. Our rationale is that an intermediate, connective filter is a strong potential reason for a user to have performed interim pivots that lead back to the same category, and its presence is very suggestive that it may indeed be what the user was thinking. In the event that leaving the filter is an error, systems should always have a mechanism for users to understand and correct where this heuristic fails.

\subsubsection{Returning Without Intermediate Filters}
In contrast, where no filters were enacted during interim pivots, we assume that the user intends to fan-out. Our rationale here is that, were the direct filter to be retained, the user will almost always arrive at exactly the same set that they started with---in our example, $A_0' = A_1'$, rendering the interim pivots meaningless.

It is possible for a subtle difference to exist without intermediate filters---for example, if an actor in $A_0'$ only acted in one movie in $M_0$ that did not have any connected directors in $D$ in the database, then that actor would be missing in the resulting set of actors $A_1'$. However, we expect that corner cases such are rare. Furthermore, there is a straightforward interpretation of a series of unfiltered pivots, that implies ever-widening sets of nodes. In the above example, without an intermediate filter on directors $D$, $A_1$ is the full set of actors that also worked with directors that worked with the original set $A_0$.

Consequently, the heuristic for intermediate pivots without filters is to remove the original direct filter upon return. Although we suspect the likelihood of errors in this case to be lower, systems that automatically remove filters should make their actions clear, and easy to revert.

\subsection{Implications For Learning From Pivots}
\label{sec:adaptiveAbstraction}
In addition to their potential in helping users more freely navigate graphs, pivots also present opportunities for system designers to develop adaptive data abstractions. As we have mentioned, a critical difficulty in visualization design is the inability to validate the accuracy of data and task abstractions before implementing a system~\cite{munzner_nested_2009}. A visualization designer must arbitrarily decide the structure of the data before implementing a visualization---all too frequently, system designers choose an abstraction that does not correctly anticipate users' tasks or data, only to discover this error after significant work has been put into implementing a system.

Exposing users to purely structural operations like the graph pivot can make these misunderstandings more apparent; we saw examples of this in our study. Users were often not aware that they had successfully completed Task 1, they would often go to the ``wrong'' part of the interface to filter a set of nodes, and they would often type node classes in the search box, such as ``Team,'' instead of querying node attributes. While this behavior may have been in part due to their unfamiliarity with the interface, it makes sense that users would not immediately know whether to think of a value as an attribute of a node, a distinct node entity, an edge, or even a node class. These are arbitrary decisions that may or may not correspond to the user's expectations.

Pivots do not merely expose the arbitrary nature of certain data abstraction decisions. They can also work the other way, in that they expose what a user expects the data abstraction to be. The abstraction-agnostic nature of pivots presents an opportunity to learn about and adapt to the semantics of the data on the fly, rather than having to anticipate it completely from the start. A series of
pivots is a very simple---yet explicit---indication of the data semantics from the user's perspective. When a series of pivots is unambiguous, it creates an unprecedented theoretical possibility: a system could observe user behavior, and reshape the data on the fly to more appropriately match the users' tasks and data.

\subsubsection{Adaptive Connections}
\begin{figure*}[t!]
\begin{center}

\includegraphics[width=\linewidth]{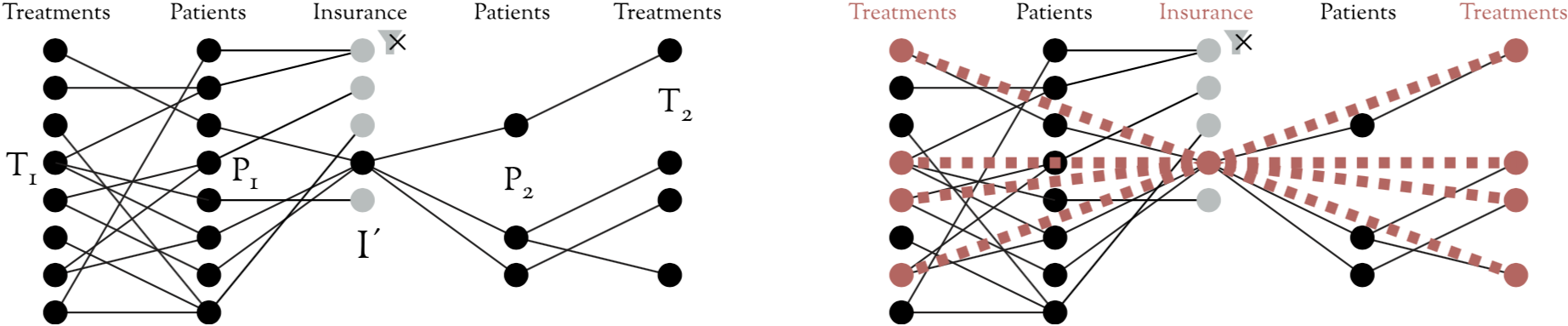}

\caption{In this scenario, doctors frequently perform connective filtering on potential treatments by the insurance companies that have covered those treatments for patients in the past. The system observes this behavior, and adapts the underlying data abstraction in response, adding direct connections between treatments and insurance companies through patients.}
\vspace{-0.6cm}
\label{fig:newEdges}
\end{center}
\end{figure*}

For example, suppose in the hospital database scenario in Figure~\ref{fig:newEdges}, that doctors must frequently determine which treatments can be prescribed based on a patient's insurance provider. However, let us assume that in the initial graph abstraction, insurance providers and treatments are only connected through patients. While pivoting and filtering make it possible to identify which treatments specific insurance companies have allowed, this is a very roundabout way of answering that question, and it encounters the somewhat complex semantics of connective filtering that we discuss above.

In this example, the system could observe users performing frequent pivots from treatments $T_0$, to patients $P_0$, to insurance providers $I$, applying a filter $I'$, and pivoting back ($P_1, T_1$). When this pattern reaches a certain threshold of usage, the system could automatically add a set of edges that directly connect the insurance providers with the prescribed treatments, bypassing the need to pivot through patients. From usage patterns alone, a machine could automatically ``invent'' a new category of semantically meaningful edges, in this case, edges that indicate that a specific insurance company has covered a specific treatment in the past. These new edges would allow users to move directly between these elements and make correlations without having to pivot through the patient nodes---enhancing both the semantic relevance of the underlying data abstraction, as well as database efficiency.

\subsubsection{Adaptive Attributes}
Edge topology is not the only arbitrary schema decision a technical expert may make with regard to a graph data abstraction. For example, as we have discussed above, the decision whether something is a node or an attribute of a node is arbitrary, and may or may not be amenable to a user's task. These decisions, too, can benefit from observing user behavior in the context of a series of pivots.

Suppose that administrators at a university are frequently trying to pair students with professors from their home country. Let us assume that in the initial data system, the home countries of both students and professors are stored as an attribute of those nodes.

The system could observe users frequently using this attribute to correlate these two types of nodes. In response, the system can push the country attribute of student and professor nodes out into the graph as independent country nodes, allowing users to make direct pivots between students and professors from the same country.

\subsubsection{Advantages and Limitations of Learning From Pivots}
The result of these alterations to the underlying data structure is that the graph can adapt to better support current and new questions. The system learns which connections hold the most valuable, real-world knowledge and exposes those connections as directly as possible. These updates can be performed automatically, either as the relevant patterns are detected, or as the processing and storage resources become available to support the added complexity. Overall, this kind of system would allow the underlying data abstraction to be improved \textit{in situ}, without constant collaboration between the technical experts and the domain experts. Using this method of back-filling the database structure, the graph automatically adapts to be able to efficiently deliver what users need from it.

Of course, the broad decision to interpret the data as a graph is still an arbitrary, \textit{a priori} assumption that a technical expert makes that they can not validate without implementing and evaluating a system with user testing. Learning from graph pivots only provides some wiggle room within that broad decision---the pitfall of choosing the wrong broad data abstraction remains.

Furthermore, a visualization that relies on an adaptive data structure must be somewhat general, like Jacob's Ladder, employing general techniques such as graph pivots. Specialized visualizations that rely on dataset and domain-specific semantics, such as anticipating certain entities as nodes, and others as node attributes, will not be able to make use of this kind of approach.

While our experience with Jacob's Ladder has exposed these two examples---adaptive connections and adaptive attributes---as ways that graphs could self-adapt to changing user needs, we can not enumerate all the possibilities for self-adapting data abstractions. Instead, by introducing the theoretical possibility of adaptive graph data abstractions, we advocate for future work into similar approaches for graphs and other data abstraction types. It may be possible, for example, for a system to automatically derive new set definitions as users interact with general-purpose set visualization systems such as UpSet~\cite{lex_upset:_2014}, or to automatically pre-compute frequent weighted attribute combinations in general-purpose ranking systems such as LineUp~\cite{gratzl_lineup:_2013}.

\section{Conclusions and Future Work}
Our purpose in this work has been to articulate how users understand pivots, how they can be useful, and to explore a visually scalable technique for representing pivots---however, in our efforts to describe pivots in a general task sense, agnostic to any particular graph's schema, size, or complexity, we do not discuss how to compute pivots efficiently. In a computational sense, however, pivots are not agnostic to schema, size, nor complexity, and we leave computational scalability challenges for future work.

Across our lab tests and user tests, Jacob's Ladder helped us to examine the expressive abilities and ambiguities that arise when constructing queries using sequences of pivot operations. The graph pivot is a very simple and intuitive, yet powerful operation that shows promise for the future of graph data analysis, especially as it does not suffer from visual scalability with respect to the size of a graph. When coupled with filtering, users with a diverse range of expertise were able to discover and extract data subsets of interest at this aggregate, categorical level.

Although we have demonstrated that visualizing local topology is not necessary for many analysis tasks, our observations suggested that an even higher-level overview of the global schema would be beneficial to help users plan where to filter or pivot. In continuing this work, we plan to more thoroughly test smart pivoting heuristics; build and test systems that adapt their abstractions; and further explore computational scalability issues.

Finally, our tests have exposed, but not fully answered, two important questions relating to pivots: whether smart pivots can accurately predict user intent with respect to filters, and how the simple nature of the graph pivot could make it possible to learn semantic information from user behavior, potentially granting visualization designers some flexibility in their initial data abstractions. Future systems that adapt their underlying data structure to user queries should become more semantically relevant.


\bibliographystyle{abbrv}
\bibliography{jacobsLadder}

\begin{thebibliography}{10}

\bibitem{auber_tulip_2004}
D.~Auber.
\newblock Tulip \textemdash{} {{A Huge Graph Visualization Framework}}.
\newblock In M.~J\"unger and P.~Mutzel, editors, {\em Graph {{Drawing
  Software}}}, Mathematics and {{Visualization}}, pages 105--126. {Springer
  Berlin Heidelberg}, Berlin, Heidelberg, 2004.

\bibitem{bastian_gephi_2009}
M.~Bastian, S.~Heymann, and M.~Jacomy.
\newblock Gephi : {{An Open Source Software}} for {{Explorating}} and
  {{Manipulating Networks}}.
\newblock {\em Proceedings of the Third International ICWSM Conference},
  page~2, 2009.

\bibitem{bezerianos_graphdice:_2010}
A.~Bezerianos, F.~Chevalier, P.~Dragicevic, N.~Elmqvist, and J.~D. Fekete.
\newblock Graphdice: {{A System}} for {{Exploring Multivariate Social
  Networks}}.
\newblock In {\em Proceedings of the 12th {{Eurographics}} / {{IEEE}} - {{VGTC
  Conference}} on {{Visualization}}}, {{EuroVis}}'10, pages 863--872,
  Chichester, UK, 2010. {The Eurographs Association \& John Wiley \& Sons,
  Ltd.}

\bibitem{bonsignore_first_2009}
E.~M. Bonsignore, C.~Dunne, D.~Rotman, M.~Smith, T.~Capone, D.~L. Hansen, and
  B.~Shneiderman.
\newblock First {{Steps}} to {{Netviz Nirvana}}: {{Evaluating Social Network
  Analysis}} with {{NodeXL}}.
\newblock In {\em 2009 {{International Conference}} on {{Computational
  Science}} and {{Engineering}}}, pages 332--339, Vancouver, BC, Canada, 2009.
  {IEEE}.

\bibitem{cao_g-miner:_2015}
N.~Cao, Y.-R. Lin, L.~Li, and H.~Tong.
\newblock G-{{Miner}}: {{Interactive Visual Group Mining}} on {{Multivariate
  Graphs}}.
\newblock In {\em Proceedings of the 33rd {{Annual ACM Conference}} on {{Human
  Factors}} in {{Computing Systems}}}, {{CHI}} '15, pages 279--288, New York,
  NY, USA, 2015. {ACM}.

\bibitem{dork_pivotpaths:_2012}
M.~Dork, N.~H. Riche, G.~Ramos, and S.~Dumais.
\newblock {{PivotPaths}}: {{Strolling}} through {{Faceted Information Spaces}}.
\newblock {\em IEEE Transactions on Visualization and Computer Graphics},
  18(12):2709--2718, Dec. 2012.

\bibitem{dunne_graphtrail:_2012}
C.~Dunne, N.~Henry~Riche, B.~Lee, R.~Metoyer, and G.~Robertson.
\newblock {{GraphTrail}}: {{Analyzing Large Multivariate}}, {{Heterogeneous
  Networks While Supporting Exploration History}}.
\newblock In {\em Proceedings of the {{SIGCHI Conference}} on {{Human Factors}}
  in {{Computing Systems}}}, {{CHI}} '12, pages 1663--1672, New York, NY, USA,
  2012. {ACM}.

\bibitem{ferrucci_ibm's_2011}
D.~A. Ferrucci.
\newblock {{IBM}}'s {{Watson}}/{{DeepQA}}.
\newblock In {\em Proceedings of the 38th {{Annual International Symposium}} on
  {{Computer Architecture}}}, {{ISCA}} '11, pages~--, New York, NY, USA, 2011.
  {ACM}.

\bibitem{ghani_multinode-explorer:_2012}
S.~Ghani, N.~Elmqvist, and D.~S. Ebert.
\newblock {\em {{MultiNode}}-{{Explorer}}: {{A Visual Analytics Framework}} for
  {{Generating Web}}-Based {{Multimodal Graph Visualizations}}}.
\newblock {The Eurographics Association}, 2012.

\bibitem{gratzl_lineup:_2013}
S.~Gratzl, A.~Lex, N.~Gehlenborg, H.~Pfister, and M.~Streit.
\newblock {{LineUp}}: {{Visual Analysis}} of {{Multi}}-{{Attribute Rankings}}.
\newblock {\em IEEE Transactions on Visualization and Computer Graphics},
  19(12):2277--2286, Dec. 2013.

\bibitem{greenberg_usability_2008}
S.~Greenberg and B.~Buxton.
\newblock Usability {{Evaluation Considered Harmful}} ({{Some}} of the
  {{Time}}).
\newblock In {\em Proceedings of the {{SIGCHI Conference}} on {{Human Factors}}
  in {{Computing Systems}}}, {{CHI}} '08, pages 111--120, New York, NY, USA,
  2008. {ACM}.

\bibitem{heer_orion:_2011}
J.~Heer and A.~Perer.
\newblock Orion: {{A System}} for {{Modeling}}, {{Transformation}} and
  {{Visualization}} of {{Multidimensional Heterogeneous Networks}}.
\newblock In {\em {{IEEE Visual Analytics Science}} $\backslash$\&
  {{Technology}} ({{VAST}})}, page~10, 2011.

\bibitem{kang_exploring_2006}
H.~Kang, C.~Plaisant, B.~Lee, and B.~B. Bederson.
\newblock Exploring {{Content}}-actor {{Paired Network Data Using Iterative
  Query Refinement}} with {{NetLens}}.
\newblock In {\em Proceedings of the 6th {{ACM}}/{{IEEE}}-{{CS Joint
  Conference}} on {{Digital Libraries}}}, {{JCDL}} '06, pages 372--372, New
  York, NY, USA, 2006. {ACM}.

\bibitem{lee_task_2006}
B.~Lee, C.~Plaisant, C.~S. Parr, J.-D. Fekete, and N.~Henry.
\newblock Task {{Taxonomy}} for {{Graph Visualization}}.
\newblock In {\em Proceedings of the 2006 {{AVI Workshop}} on {{BEyond Time}}
  and {{Errors}}: {{Novel Evaluation Methods}} for {{Information
  Visualization}}}, {{BELIV}} '06, pages 1--5, New York, NY, USA, 2006. {ACM}.

\bibitem{lex_upset:_2014}
A.~Lex, N.~Gehlenborg, H.~Strobelt, R.~Vuillemot, and H.~Pfister.
\newblock {{UpSet}}: {{Visualization}} of {{Intersecting Sets}}.
\newblock {\em IEEE Transactions on Visualization and Computer Graphics},
  20(12):1983--1992, Dec. 2014.

\bibitem{liu_ploceus:_2014}
Z.~Liu, S.~B. Navathe, and J.~T. Stasko.
\newblock Ploceus: {{Modeling}}, visualizing, and analyzing tabular data as
  networks.
\newblock {\em Information Visualization}, 13(1):59--89, Jan. 2014.

\bibitem{munzner_nested_2009}
T.~Munzner.
\newblock A {{Nested Model}} for {{Visualization Design}} and {{Validation}}.
\newblock {\em IEEE Transactions on Visualization and Computer Graphics},
  15(6):921--928, Nov. 2009.

\bibitem{nielsen_abyss-explorer:_2009}
C.~Nielsen, S.~Jackman, I.~Birol, and S.~Jones.
\newblock {{ABySS}}-{{Explorer}}: {{Visualizing Genome Sequence Assemblies}}.
\newblock {\em IEEE Transactions on Visualization and Computer Graphics},
  15(6):881--888, Nov. 2009.

\bibitem{renoust_detangler:_2015}
B.~Renoust, G.~Melan{\c c}on, and T.~Munzner.
\newblock Detangler: {{Visual Analytics}} for {{Multiplex Networks}}.
\newblock {\em Computer Graphics Forum}, 34(3):321--330, 2015.

\bibitem{shannon_cytoscape:_2003}
P.~Shannon, A.~Markiel, O.~Ozier, N.~S. Baliga, J.~T. Wang, D.~Ramage, N.~Amin,
  B.~Schwikowski, and T.~Ideker.
\newblock Cytoscape: {{A Software Environment}} for {{Integrated Models}} of
  {{Biomolecular Interaction Networks}}.
\newblock {\em Genome Res.}, 13(11):2498--2504, Jan. 2003.

\bibitem{singhal_introducing_2012}
A.~Singhal.
\newblock Introducing the {{Knowledge Graph}}: Things, not strings.
\newblock
  https://www.blog.google/products/search/introducing-knowledge-graph-things-not/,
  May 2012.

\bibitem{stasko_jigsaw:_2007}
J.~Stasko, C.~Gorg, Z.~Liu, and K.~Singhal.
\newblock Jigsaw: {{Supporting Investigative Analysis}} through {{Interactive
  Visualization}}.
\newblock In {\em 2007 {{IEEE Symposium}} on {{Visual Analytics Science}} and
  {{Technology}}}, pages 131--138, Sacramento, CA, USA, Oct. 2007. {IEEE}.

\bibitem{vandenelzen_multivariate_2014}
S.~{van den Elzen} and J.~J. {van Wijk}.
\newblock Multivariate {{Network Exploration}} and {{Presentation}}: {{From
  Detail}} to {{Overview}} via {{Selections}} and {{Aggregations}}.
\newblock {\em IEEE Transactions on Visualization and Computer Graphics},
  20(12):2310--2319, Dec. 2014.

\bibitem{vanham_"search_2009}
F.~{van Ham} and A.~Perer.
\newblock ``{{Search}}, {{Show Context}}, {{Expand}} on {{Demand}}'':
  {{Supporting Large Graph Exploration}} with {{Degree}}-of-{{Interest}}.
\newblock {\em IEEE Transactions on Visualization and Computer Graphics},
  15(6):953--960, Nov. 2009.

\bibitem{wattenberg_visual_2006}
M.~Wattenberg.
\newblock Visual {{Exploration}} of {{Multivariate Graphs}}.
\newblock In {\em Proceedings of the {{SIGCHI Conference}} on {{Human Factors}}
  in {{Computing Systems}}}, {{CHI}} '06, pages 811--819, New York, NY, USA,
  2006. {ACM}.

\end{thebibliography}
\end{document}